\documentclass[twocolumn]{aastex631}

\usepackage{amsmath}
\usepackage{booktabs} 
\usepackage{multirow}
\usepackage{pifont}

\makeatletter
\let\tablenum\@undefined
\makeatother
\usepackage{siunitx}

\definecolor{darkgreen}{RGB}{0,100,0}    
\definecolor{darkred}{RGB}{139,0,0}      
\definecolor{yellow1}{RGB}{255,223,10}

\newcommand{\cmark}{\textcolor{darkgreen}{\checkmark}}
\newcommand{\xmark}{{\color{darkred}\ding{55}}}

\newcommand{\statusbox}[2]{%
  \protect\colorbox{#1}{\makebox[1.8em][c]{\textcolor{white}{\textbf{#2}}}}%
}

\usepackage{tikz}

\usepackage{verbatim}

\shorttitle{}
\shortauthors{}

\graphicspath{{./}{figures/}}

\begin{document}

\title{The Goldilocks problem for detecting water in terrestrial planets: Constraining water abundances in the mid-IR with LIFE} 

\correspondingauthor{Rugheimer, S.}
\email{s.rugheimer@ed.ac.uk}

\author{S. Rugheimer}
\affiliation{School of Physics and Astronomy, University of Edinburgh Blackford Hill, Edinburgh EH9 3HJ, UK}
\affiliation{Department of Physics and Astronomy, York University, 4700 Keele St., Toronto, ON M3J 1P3, Canada}

\author{E. Alei}
\affiliation{ETH Zurich, Institute for Particle Physics \& Astrophysics, Wolfgang-Pauli-Str. 27, 8093 Zurich, Switzerland}
\affiliation{NPP Fellow, NASA Goddard Space Flight Center, 8800 Greenbelt Rd., 20771 Greenbelt, MD, USA }

\author{B.S. Konrad}
\affiliation{ETH Zurich, Institute for Particle Physics \& Astrophysics, Wolfgang-Pauli-Str. 27, 8093 Zurich, Switzerland}

\author{B. Taysum}
\affiliation{Institute for Space Research, German Aerospace Center (DLR), Rutherfordstraße 2, Berlin, 12489, Germany}

\author{J. L. Grenfell}
\affiliation{Institute for Space Research, German Aerospace Center (DLR), Rutherfordstraße 2, Berlin, 12489, Germany}
\affiliation{Institute for Geological Sciences, Free University of Berlin, Malteserstraße 74-100,
12249 Berlin, Germany}

\author{T. Lichtenberg}
\affiliation{Kapteyn Astronomical Institute of the University of Groningen, Landleven 12, 9747 AD Groningen, The Netherlands}

\author{D. Kitzmann}
\affiliation{Space Research and Planetary Sciences, Physics Institute, University of Bern, Gesellschaftsstrasse 6, 3012 Bern, Switzerland}
\affiliation{Center for Space and Habitability, University of Bern, Gesellschaftsstrasse 6, 3012 Bern, Switzerland}

\author{F. van der Tak}
\affiliation{Kapteyn Astronomical Institute of the University of Groningen, Landleven 12, 9747 AD Groningen, The Netherlands}
\affiliation{Space Research Organisation Netherlands (SRON), Groningen, The Netherlands}

\author{S.P. Quanz}
\affiliation{ETH Zurich, Institute for Particle Physics \& Astrophysics, Wolfgang-Pauli-Str. 27, 8093 Zurich, Switzerland}
\affiliation{ETH Zurich, Department of Earth and Planetary Sciences, Sonneggstrasse 5, 8092 Zurich, Switzerland}

\author{LIFE collaboration}

\begin{abstract}

We investigate how well the Large Interferometer for Exoplanets (LIFE) mission concept can detect habitable conditions on exoplanets through the presence of atmospheric water vapor as a proxy for surface oceans. We model the atmosphere of a pre-biotic Earth-like planet across a range of water concentrations, from water-poor to water-rich, with surface partial pressures from 10$^{-7}$ to 1 bar of H$_2$O. We simulate LIFE-like noise at spectral resolutions R = 50 and 100 using \textsc{LIFEsim} and perform Bayesian atmospheric retrievals to determine the technical requirements for LIFE to confirm habitability. We model three vertical water distributions: a vertically constant profile, a Manabe–Wetherald based Earth-like profile, and a diffusion and photochemistry profile to test how the assumed vertical structure influences the retrieved abundances. Clouds are not modeled. We find the ability for LIFE to detect water strongly depends on the vertical profile assumed. LIFE is unable to constrain the highest water cases and provides upper limits on low water planets. For the highest water abundances, absorption features saturate and reduce sensitivity to characterize precise H$_2$O levels. Water vapor is not detectable in any profile modeled for $\leq10^{-6}$ bar in surface water, comparable to Mars. For an Earth-like profile, LIFE could constrain H$_2$O concentrations from $\sim10^{-3}$ to 1 bar, spanning below and above present-day Earth concentrations of 10$^{-2}$ bar. Detectable atmospheric water may imply surface oceans, as water is highly reactive and rapidly removed by surface mineral reactions. Thus, LIFE can characterize water abundances indicative of habitable surface conditions.

\end{abstract}


\section{Introduction} \label{sec:intro}

We are now entering the era of detecting and attempting to characterize rocky planet atmospheres with JWST \citep[see e.g.][]{moran2023, greene2023, zieba2023, bello2025, ducrot2025} with large ground-based observatories such as the ELT coming online later in this decade \citep[][and references therein]{padovani2023}. However, the number of potentially habitable planets in these initial searches with JWST will be limited to a handful of targets because JWST has only access to known transiting rocky exoplanets where the achievable SNR is comparatively low.  \citep{deming2009, kaltenegger2009}. The next generation of flagship missions such as the Large Interferometer for Exoplanets (LIFE) \citep{quanz2022a} and the Habitable Worlds Observatory (HWO) \citep{national2021, feinberg2024} will leverage direct detection techniques to assess the habitability of dozens of exoplanetary systems. 

Due to their different wavelength regimes, LIFE and HWO are complementary missions with HWO observing reflected starlight in the ultraviolet/visible/near-infrared (UV/VIS/near-IR) and LIFE observing thermal emission in the mid-infrared (mid-IR) \citep[see e.g.][]{alei2024}. These or other comparably sized missions are required to characterize a statistically significant sample size of habitable planets where even a null result will provide a meaningful constraint on the abundance of life in the Universe \citep{quanz2022b, angerhausen2025}. 

The Large Interferometer for Exoplanets (LIFE) will aim to characterize about 30-50 habitable planets around Sun-like F, G, K and a slightly smaller number orbiting M stars. Using nulling interferometry in the mid-IR to suppress starlight, LIFE would allow us to constrain the radius, atmospheric pressure, and surface temperature of terrestrial exoplanets, provide unique information about their atmospheric structure and composition, and search for biosignatures and even technosignatures in their atmospheres \citep{quanz2018, quanz2022a, konrad2022LIFEIII,alei2022LIFEV, Konrad2023, schwieterman2024, mettler2024, angerhausen2024}.

One of the first tasks of these future missions will be to assess a target not just for its habitability potential, but indeed if it currently has liquid water and is habitable for life as we know it. This will aid in target selection to prioritize planets for further follow-up. Although it is unknown whether water is a necessary condition for life and biochemistry in general, all life on Earth uses water for its solvent \cite[see][for a review]{westall2018}. Thus, it will remain a key target when searching for the first remote biosignatures as it indicates habitability and a temperate environment, which will provide vital context for interpreting carbon-based biochemistry biosignatures \citep[e.g.][]{rauer2011, kaltenegger2012, seager2016, meadows2018, wogan2020, quanz2022a}. Due to the centrality of water to life as we know it, NASA uses a ``follow the water'' approach to seeking out habitable environments in the Solar System \citep{hendrix2019}, and the Habitable Zone (HZ) is commonly defined as the region around a star where liquid water can be stable on the surface of an exoplanet \citep{kastingHZ1993, Kopparapu2013}. 

One previously explored way of directly detecting an ocean on an exoplanet is to look for ocean glint \citep{sagan1993, robinson2010, lustig2018, cowan2025}. Ocean glint can be detected by a sharp increase in reflectance at the crescent phases due to the specular reflection of water at oblique angles relative to the illumination source and the degree of polarization \citep{mccullough2006,zugger2010, trees2022}. However, ocean glint may be difficult to detect due to clouds and because it is visible only at crescent phases $>$120$^\circ$, which might necessitate a small inner working angle (IWA) \citep{cowan2025}. Ocean glint is unable to distinguish an Earth-like ocean coverage from a full water world planet unless land can also be detected \citep{ ulses2025}. Ocean glint in theory is detectable in the VIS and near-IR range, though a detection is not trivial because of the dependence of the IWA on wavelength \citep{cowan2025}. Another way to directly detect oceans in the visible is through spatial mapping of the dark oceans compared to the relatively bright continents \citep{cowan2009} and potentially through detecting scattering of water clouds at gibbous phases \citep{vaughan2023}. 

Direct detection of oceans via glint or spatial mapping is not detectable with LIFE in the mid-IR. However, in the infrared, there are many strong water features that indicate atmospheric water vapor \citep[see e.g.][]{desmarais2002}. LIFE might be able to directly constrain surface pressure and temperature \citep{konrad2022LIFEIII} which are vital for understanding planetary habitability and context and are more difficult to constrain in the visible \citep{alei2024}. Additionally, while there are strong water features throughout the spectrum, the abundance is harder to quantify in the VIS/NIR due to stronger degeneracies with clouds \citep{feng2018, damiano2022} and surface albedo MIR lines more directly trace the water column abundance \citep{konrad2024}.

In this work, we assess the ability for the LIFE mission concept to characterize the water abundance and current habitability of an Earth-like exoplanet. We consider three  water profiles for terrestrial planets with varying surface water concentrations. We then run noise estimates with \textsc{LIFEsim} and perform Bayesian atmospheric retrievals on the resulting simulated spectra of these habitable planets to assess the sensitivity of LIFE to water abundance.

In Section 2, we describe our model and simulations set-up and in Section 3 we report the climate and simulated observation results of an Earth-like planet of varying water concentrations. In Sections 4 and 5 we summarize the results and discuss their implications.

\section{Methods} \label{sec:methods}

\subsection{Model description}

For modeling the temperature, pressure, and water profiles in our vertically constant water profile and scaled Earth-like profile we implement a 1D climate model for high CO$_2$, non-oxygenated terrestrial atmospheres with 100 layers \citep[see][and references therein]{kasting1984,pavlov2001,kharecha2005,segura2007, rugheimer2018}. We set the solar zenith angle to 60$^\circ$ to represent the global average treating the planet as a Lambertian sphere. The net absorbed stellar radiation is calculated by a delta two-stream approximation \citep{toon1989}, and correlated-k coefficients parameterize the absorption by H$_2$O, CO$_2$, and CH$_4$ \citep{pavlov2000}. When modeling the diffusion and photochemistry profile, we implement a coupled 1D climate and photochemical network, \texttt{1D-TERRA} with 1127 reactions for 128 species originating from \citet{pavlov2002} and updated and described in \citet{wunderlich2020, wunderlich2021}.

In most 1D photochemical models the vertical transport is approximated by eddy diffusion. How much water is transported into the stratosphere depends on the presence of a cold trap and the assumptions of the eddy diffusion coefficients which vary by several orders of magnitude for the terrestrial planets in our Solar System \citep{massie1981, krasnopolsky2010,krasnopolsky2012, wunderlich2020}. We use eddy diffusion coefficients calculated based on the considerations of temperature and atmospheric composition profiles using the methodology and formulation in \citet{Gierasch1985} and \citet{gao2015}.

Clouds are not self-consistently calculated in the climate model. The climatic effects of clouds on the temperature/pressure profile are included by increasing the surface albedo of the modern Earth-Sun system to 0.20 from the measured surface albedo of 0.12 \citep{trenberth2009} to maintain an Earth-like surface temperature of 290 K \citep{[following][] segura2003, segura2005,rugheimer2013}. 

We model the mid-IR thermal emission spectrum with a 1D radiative transfer code, \texttt{petitRADTRANS} \citep{molliere2019,molliere2020,alei2022LIFEV} at spectral resolutions, R = $\lambda / \Delta \lambda$ of 50 and 100. These resolutions are chosen since R=50 was the minimum mission requirement in \citep{konrad2022LIFEIII}, yet recent work demonstrated that LIFE would need to have an R=100 to accurately constrain methane \citep{konrad2024}. The emission spectrum is calculated from the dominant spectroscopically active molecules in a prebiotic, terrestrial planet, namely CO$_2$, H$_2$O, CH$_4$, and N$_2$, including collision induced absorption. For low resolution spectra, \texttt{petitRADTRANS} calculates opacities using the correlated-k method \citep{goody1989,lacis1991,fu1992}.

The calculated mid-IR spectra are passed to \textsc{LIFEsim} \citep{dannert2022LIFEII} to simulate realistic noise from the LIFE mission concept. \textsc{LIFEsim} estimates wavelength dependent S/N for all major astrophysical noise sources including stellar leakage, local zodiacal dust emission, and exo-zodiacal dust emission at three times the local zodiacal level \citep{dannert2022LIFEII}. We assume the observations are limited by astrophysical noise, thereby treating the instrumental stability as a derived requirement rather than a fixed parameter \citep[see][for a first discussion on stability limits]{dannert2022LIFEII, dannert2025thesis}. Detailed modeling of instrumental systematic effects and their trade-offs with integration times is the subject of ongoing work \citep{huber2024, huber2025, dannert2025} and is outside the scope of this work.

We then use the retrieval framework developed for other works within the LIFE project\footnote{\url{https://github.com/LIFE-SpaceMission/LIFE-Retrieval-Framework}} \citep{konrad2022LIFEIII,alei2022LIFEV} to retrieve constraints for the atmospheric profile and abundances for our model planets. The retrieval framework combines \texttt{pyMultiNest} \citep{buchner2014}, a nested sampling approach \citep{skilling2006} for Bayesian parameter inference, with \texttt{petitRADTRANS} \citep{molliere2019,molliere2020,alei2022LIFEV}, assumes vertically constant abundance profiles for all species, and parametrizes the thermal profile with a fourth-order polynomial \citep[following][]{konrad2022LIFEIII,alei2022LIFEV, Konrad2023}.

\subsection{Simulation set-up}\label{simsetup}

We model an early-Earth analogue planet with a prebiotic atmosphere dominated by CO$_2$ and N$_2$ orbiting a young solar analogue at 3.9 Ga from \citet{claire2012} at a distance of 10pc. We initially varied H$_2$O concentrations from $10^{-15}$ bar to 1 bar to span the conditions of a dry planet to a water-rich atmosphere. We assume a total surface pressure of two bar with 0.2 bar CO$_2$, a mixing ratio of CH$_4$ of $1.65 \times 10^{-6}$, and with N$_2$ as filling gas. The top of the atmosphere pressure is set to 1$\times 10^{-6}$ bar. 

In our forward model, we assume constant vertical abundance profiles for N$_2$, CO$_2$ and CH$_4$. For H$_2$O we model three different profiles to determine the impact of the shape of profile with altitude on the retrieved abundance and sensitivity of LIFE to constrain water in different circumstances. 

\begin{figure*}[ht!]
\centering
\includegraphics[width=0.7\textwidth,angle=0]{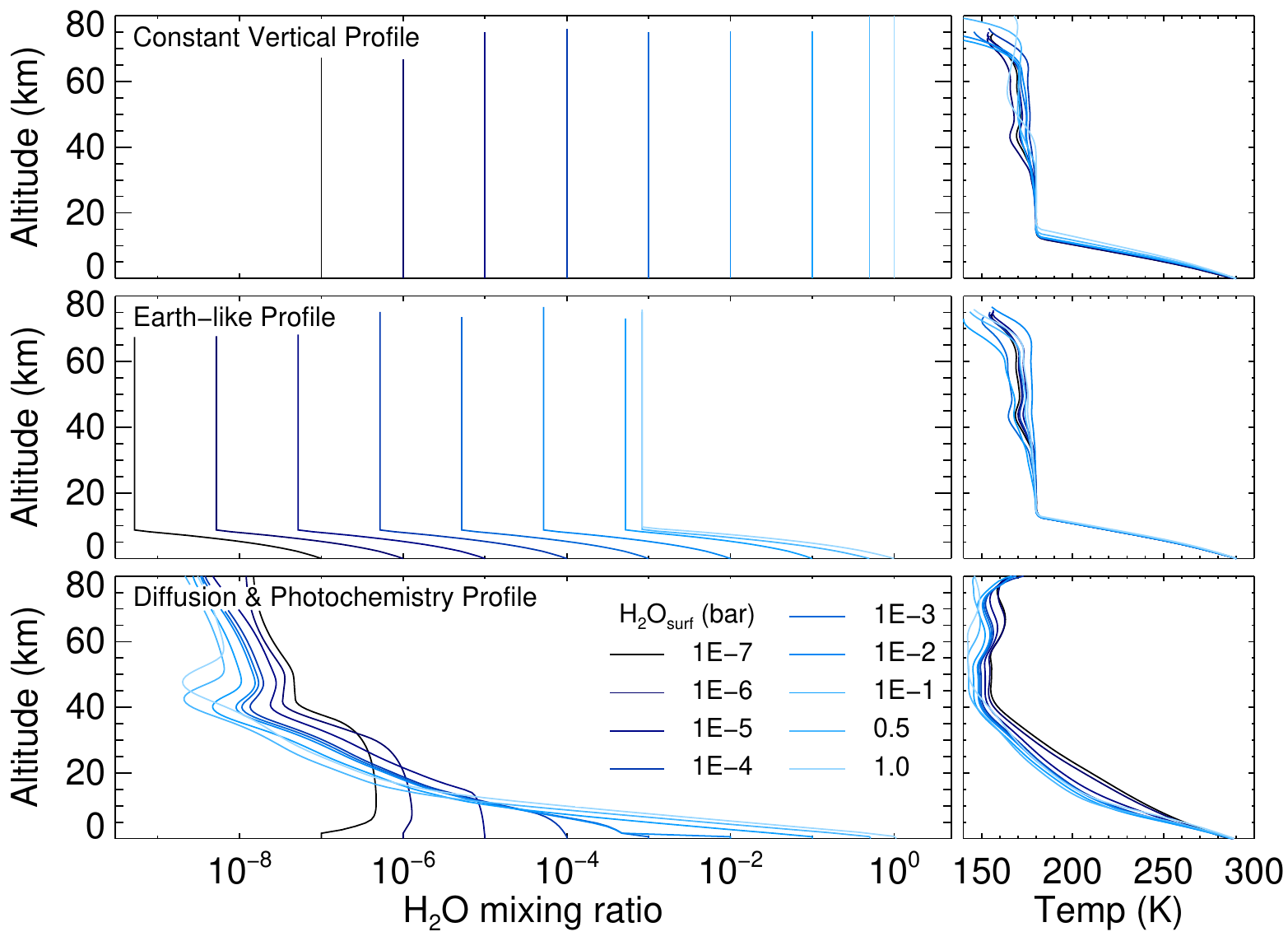}
\caption{Water mixing ratios vs. altitude for each of our three profile assumptions (left) and the temperature vs altitude profile (right). The surface mixing ratio of water is fixed, spanning concentrations from $10^{-7}$ to 1 bar. The water profile above the surface is modeled to be constant with height (top panel), an Earth-like profile (middle panel), or from only diffusion and photochemical production (bottom panel).  \label{h2omixing}}
\end{figure*}

The first water profile considered is an altitude-invariant water profile (top panel of Fig. \ref{h2omixing}). This profile is likely not physical for temperate rocky planets, but it is included since most retrieval models intrinsically assume vertically constant profiles for all molecular species in an atmosphere, including water. Additionally, the water profile approaches a vertically constant profile during a moist and run-away greenhouse scenario when the surface temperature rises and evaporation increases \citep[see Fig. 1 and Fig. 2 in][respectively]{kasting2015,kastingHZ1993}. 

The second water profile considered is a \citet{manabe1967} based profile similar to Earth's atmosphere (middle panel of Fig. \ref{h2omixing}). In this framework, the temperature in the troposphere decreases following the moist adiabatic lapse rate, while the stratosphere is in radiative equilibrium and is isothermal due to the lack of ozone and therefore absence of an inversion in a prebiotic atmosphere. The saturation vapor pressure exponentially decreases with temperature via the Clausius-Clapeyron relation, and so the ability for the atmosphere to retain water falls off rapidly in the troposphere. The falling saturation vapor pressure causes condensation and precipitation. In this model, we fix the relative humidity based on the surface water abundance but keep the same profile for Earth's water in the troposphere and stratosphere, leading to a temperature- and pressure-dependent H$_2$O distribution. In the stratosphere, H$_2$O abundances remain low and relatively uniform, maintained by diffusion above the tropopause.  

The third water profile considered is one where the atmospheric water composition is determined entirely by the diffusion of water from the surface, assuming a fixed relative humidity based on the surface water abundance in the surface layer, and by photochemical production in the atmosphere (see bottom panel of Fig. \ref{h2omixing}). The upper atmosphere is suppressed to a relative humidity of 10\% in the stratosphere after precipitation. The diffusion and photochemistry are free to produce lower levels of H$_2$O dynamically, but if the relative humidity exceeds 10\% the photochemistry forces H$_2$O to volume mixing ratios corresponding to 10\% relative humidity values. In this profile we note that the strongest photochemical mechanisms to generate water in the stratosphere is via methane oxidation \citep{johnston1998}. The net reaction is:
\begin{align}
\text{CH}_4 + 2\text{O}_2 &\rightarrow \text{CO}_2 + 2\text{H}_2\text{O} 
\end{align}\label{methaneoxidation}
Thus for each methane molecule oxidized, two water molecules are produced in the stratosphere. Methane is oxidized mainly by the hydroxyl radical, OH and UV photons. Oxygen in the net reaction is not from biology, unlike on modern Earth. In the pre-biotic Earth, this O$_2$ comes primarily from the photolysis of CO$_2$ in the upper atmosphere which depends on OH from photolysis of H$_2$O in the troposphere. This profile has higher stratospheric water abundances for the low surface water cases and lower stratospheric water abundances for the higher surface water cases than then Manabe-Wetherald profiles. This difference is in part due to the changing availability of UV photons to drive methane oxidation from a variable semi-major axis. 

After running a parameter sweep of water abundance from 1 to $10^{-15}$ bar we see no additional sensitivity to H$_2$O below a surface concentration of $10^{-7}$ bar for any shape of profile. Hence, we ran simulations with surface H$_2$O concentrations of: $10^{-7}$, $10^{-6}$, $10^{-5}$, $10^{-4}$, $10^{-3}$, $10^{-2}$, $10^{-1}$, 0.5, and 1 bar. For comparison, Earth and Mars have a surface volume mixing ratio water concentration of $\sim$ 1$\times$10$^{-2}$ and 10$^{-6}$ bar, respectively. These values allow us to identify the transition of when H$_2$O vapor starts contributing discernibly to the low-resolution mid-IR spectrum. 

Water is a dominant greenhouse gas, and thus as we increase the H$_2$O content, we move the planet outward to compensate for this additional heating to maintain an Earth-like surface temperature of 290 K.

We then use these 1D atmosphere models to calculate the IR spectra with R=50 and R=100 and perform atmospheric retrievals on the mock observations using a Bayesian parameter inference. We neglected the impact of clouds on the spectra. In the retrieval we assume a wavelength independent surface reflectance of 0.1, a common value for Earth-like planets \citep[following][]{alei2022LIFEV}.

\section{Results} \label{sec:results}

\begin{figure}[h!]
\centering
\includegraphics[angle=0,width=0.5\textwidth]{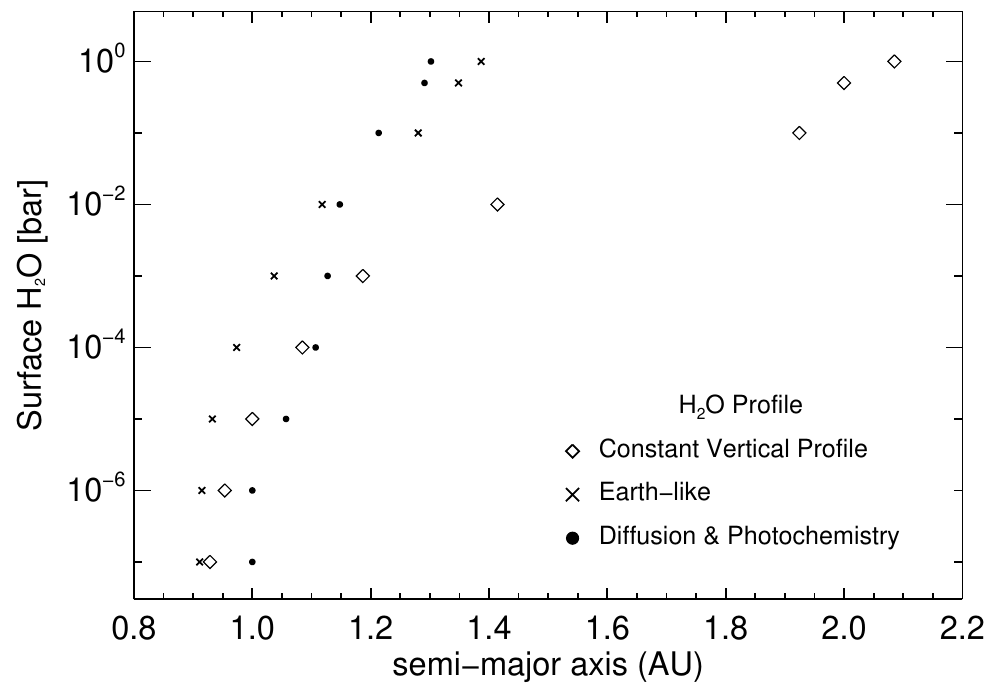}
\caption{Planet distance (AU) versus water surface mixing ratio (bar) needed to maintain an average surface temperature of 290K. While the surface mixing ratio of water is fixed, the profile higher up in the atmosphere can either be constant with height (diamond), an Earth-like (cross), or based on diffusion and photochemistry (circle). \label{distance}}
\end{figure}

We used a 1D climate model to calculate the distance a planet would need to orbit to maintain a modern Earth-like surface temperature with increasing water abundance. Fig. \ref{distance} shows the calculated semi-major axis in AU vs the surface water abundance for the three different water profiles required to maintain a surface temperature of 290K. While surface concentration is what is plotted on the y-axis, note the total water column abundance is in most cases higher for the vertically constant profile, and thus the planet distance from the host star is increased most for those models due to the increased greenhouse effect from water.

\subsection{Climate model results}

In the diffusion and photochemistry profile, for planets with smaller semi-major axes the resulting higher insolation and thus elevated UV flux causes the stratospheric CH$_4$ oxidation rate to increase, producing more stratospheric water (darker blue lines in bottom panel of Fig. \ref{h2omixing}). As the surface H$_2$O is progressively increased, a lower insolation is needed to maintain 290K, and the rate of CH$_4$ oxidation correspondingly decreases, causing the stratospheric water concentration to fall (lighter blue lines in bottom panel of Fig. \ref{h2omixing}). 

\begin{figure}
  \centering

\includegraphics[width=1.0\linewidth]{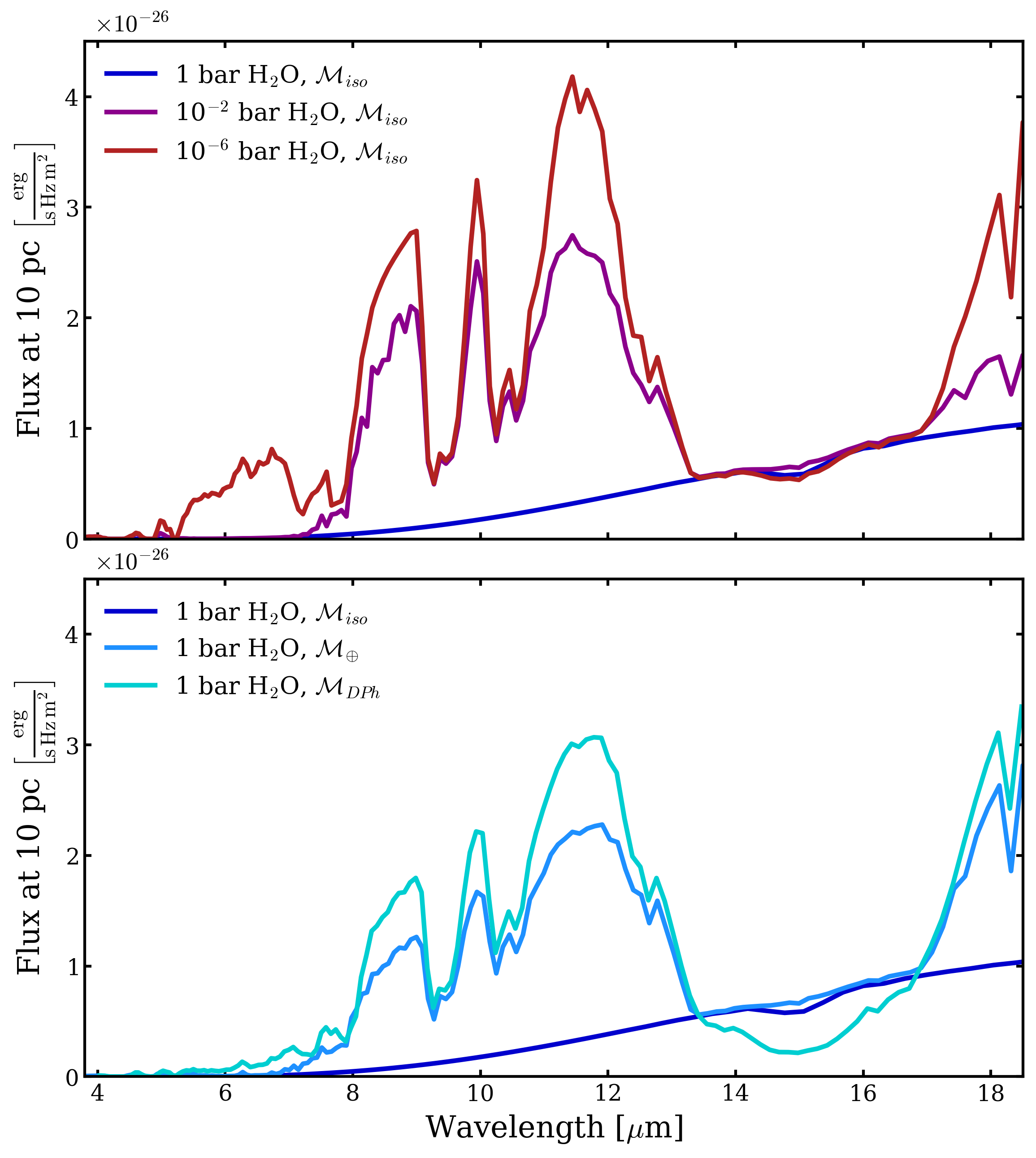}
  
  \caption{Mid-IR input spectra at R=100 calculated by \texttt{petitRADTRANS} as input to \textsc{LIFEsim} for different H$_2$O abundances and H$_2$O profiles. The top panel shows three spectra for vertically constant profile with surface H$_2$O concentrations corresponding to a water rich case of 1 bar (dark blue solid line), an Earth-like surface water case of 10$^{-2}$ bar (purple line), and a water poor, Mars-like case of 10$^{-6}$ bar (red line). The bottom panel shows a comparison between spectra with different water altitude profiles with the same surface water concentration of 1 bar for the altitude invariant profile (dark blue), the Earth-like profile (blue), and the diffusion and photochemistry profile (light blue).}
  \label{spectraIsoVariable}
\end{figure}

Figure \ref{spectraIsoVariable} shows a subset of the input spectra used in the retrieval model. The top panel shows the spectra for a low water, comparable to Mars (10$^{-6}$ bar, red), medium Earth-like water (10$^{-2}$ bar, purple), and high water (1 bar, dark blue) abundance assuming an altitude invariant water profile. Note, retrieval models typically assume that all atmospheric molecules are constant in height to reduce computational complexity and because solutions with more complex profiles are often too degenerate. 

The 1 bar vertically constant profile case has the highest total H$_2$O column abundance of any planet modeled at 50\% of the 2 bar atmosphere. Here, the H$_2$O features completely dominates the profile, and it is a near featureless spectrum apart from a slight dip at the 15 $\mu$m CO$_2$ feature (Fig. \ref{spectraIsoVariable}, dark blue line). The 1 bar H$_2$O vertically constant profile spectrum with 2 bars total pressure and a surface temperature of 290K is degenerate with a cooler blackbody with a low pressure ($<$0.1 bar), cold atmosphere ($\sim$ 180K). This is because the lower atmosphere is optically thick to IR radiation and so we are only seeing the emission from layers higher up in the atmosphere. Too much water erases its own water features as well as nearly all other features in the spectrum which is also seen in the spectra for the 0.5 and 0.1 bar vertically constant water profiles. In these cases, the actual surface temperature of the modeled planets is 290K, but the atmosphere is so opaque that the emission peaks around 28$\mu$m, and the planet looks like a colder object to a remote observer. This is similar to Venus, where the surface temperature is much hotter than one would calculate from the brightness temperature which originates from higher in the atmosphere. Comparing the 10$^{-2}$ with the 10$^{-6}$ bar model, the spectra differs in the 5-8$\mu$m and 18-20$\mu$m regions where water has strong infrared features. Both of these spectra also have strong CO$_2$ features at 9-11$\mu$m and 15$\mu$m.

The bottom panel of Fig. \ref{spectraIsoVariable} compares the spectra for the same surface water abundance but with different water altitude profiles. We show the 1 bar surface water atmospheres with a vertically constant profile (dark blue), an Earth-like based water profile (blue), and diffusion and photochemistry profile (light blue). Even with high surface concentrations, the exponential decrease in water in the troposphere for the Earth-like profile allows for our observations to penetrate much deeper into the atmosphere where there are strong H$_2$O features in the spectrum at 5-8$\mu$m and 18-20$\mu$m, as well as strong CO$_2$ features at 9-11$\mu$m and 15$\mu$m.  

\subsection{\textsc{LIFEsim} noise and retrieval results}\label{sec:retrieval}

We model observations over a wavelength range of 4-18.5$\mu$m with a fixed signal-to-noise ratio of 10 at 11.2$\mu$m\footnote{\textsc{LIFEsim} computes the noise distirbution across all wavelengths using SNR = 10 at 11.2 $\mu$m as a reference point.} and spectral resolutions of 50 and 100. We only include the R=100 runs in the figures here as modeling work has shown an advantage in that higher resolution to meet the main mission goals \citep{konrad2024}. However, we include R=50 in our analysis and in the online repository as the minimum resolution under consideration for LIFE. We then run atmospheric retrievals on these simulated observations to assess the ability of the LIFE mission to characterize this suite of Earth-sized terrestrial planets with varying atmospheric water contents and distributions. Additionally we look at the retrieved posteriors for some basic planetary parameters: surface pressure, temperature, mass, and radius. 

A summary of the retrieval results can be found in Table \ref{table_LIFESIM}. For each surface H$_2$O abundance scenario, we tabulate the model parameter $\theta_{\mathcal{M}}$, the simulation true value from the forward model, and the retrieved posteriors for each of the three water profiles: $\mathcal{M}_{\text{iso}}$, $\mathcal{M}_{\oplus}$, and $\mathcal{M}_{\text{DPh}}$ with the 1$\sigma$ confidence intervals. Following \citet{konrad2022LIFEIII}, we consider LIFE to be able to accurately retrieve a parameter if it is within $\pm$ 0.5 dex for pressure, $\pm$ 1 dex for the abundance of H$_2$O, and $\pm$ 20K for temperature (denoted by a green check or red x) compared to the true value. We classify each posterior as constrained (green), sensitivity-limit (lime), upper-limit (orange), or unconstrained (red) for each retrieved value \citep[see classification scheme details in Appendix B in][]{konrad2022LIFEIII}. 

A constrained (C) posterior indicates the posterior distribution can be approximated by a Gaussian distribution and it has strong constraints on the range of values. The mixing ratio abundances have a prior of being between 10$^{-15}$ to 1. Note, a model can have a constrained posterior but an inaccurate retrieved value. A sensitivity limit (SL) is able to exclude high but not low molecular abundances or vice versa and shows a distinct peak around the retrieved abundance but large uncertainty in one direction. An upper limit (UL) is where the retrieval can rule out any abundance higher than a threshold value but below the threshold all concentrations are equally likely as spectral features at those strengths cannot be distinguished from the noise. An unconstrained (UC) posterior is where the posterior distribution is roughly constant throughout the entire sampled prior range, showing that no further constraint over the prior range can be obtained. Because the nested sampling method under sampled the edges of the prior, we can expect the edges of the posterior to be not as well populated as the middle.


\subsubsection{Retrievals of planetary properties}
In general, we retrieve the atmospheric surface pressure to high precision (within $\pm$ 0.5 dex) except in the three highest water abundances, H$_2$O $\ge 10^{-1}$ bar in the vertically constant profile models only. The true atmospheric pressure is 2 bars, but in these three high-water constant profile cases, the spectrum is nearly flat, with only a small CO$_2$ feature at 15$\mu$m (see blue and orange lines in Fig. \ref{spectraIsoVariable}).\footnote{All corner plots and spectra are available at \linebreak \url{https://github.com/LIFE-SpaceMission/Detecting_Oceans}} In these cases, the retrieval prefers a low pressure (and low temperature) scenario rather than a high water one, and reports a high confidence as indicated by both relatively small error bars and in classifying the posteriors as constrained or a sensitivity limit. In all other runs the pressure is within $\pm$ 0.5 dex and all posteriors are constrained. 

We also retrieve the surface temperature to high precision within $\pm$ 20K in all but the high-water cases of H$_2$O $\ge 10^{-1}$ bar in the vertically constant profile and for H$_2$O $\ge 0.5$ bar in the Earth-like and diffusion/photochemistry models. In the vertically constant profiles with high water the temperature is underestimated by 107-108K, retrieving a categorically different temperature regime. In the Earth-like and diffusion/photochemistry high water cases, the temperature is underestimated by 22-35K. In these models, the retrieval still classifies the posterior as constrained and reports small error bars on the value, due to the flat spectrum and the strong H$_2$O opacity masking features from other species. 

The radius is well retrieved in all models. The retrieved radius is 0.91 to 1.05 R$_{\oplus}$ for the vertically constant profile, 0.96 to 1.04 R$_{\oplus}$ for the Earth-like, and 1.00 to 1.04 R$_{\oplus}$ for the diffusion/photochemistry models. The radius is over the true radius for the low- and medium-water cases, and under the true radius for the high-water cases. In the corner plots in Appendix \ref{sec:appendix} one can observe that the radius is correlated with temperature in the retrievals. This is because one can reduce the emitting area of the planetary surface and/or the temperature of the planet to fit a spectrum with a seemingly cooler emission temperature.

The mass is retrieved in all models, ranging from 0.8 to 1.0 M$_{\oplus}$ independent of water profile. Due to degeneracies between the molecular abundances and surface gravity (which depends on mass) the retrieval estimate is based on the assumed prior for the mass which comes from the assumed prior on the radius \citep[see][]{konrad2022LIFEIII}.

\subsubsection{Retrieval of atmospheric constituents}
Next we will cover the ability for LIFE to retrieve the main bulk constituents of our atmosphere: N$_2$, CO$_2$, and H$_2$O. Note the abundances in the Table \ref{table_LIFESIM} and in Appendix \ref{sec:appendix} are given in volumetric mixing ratios. 

We consider a species well constrained in the abundance if the retrieved value lies within $\pm$ 1 dex of the true value. Both N$_2$ and CO$_2$ are well mixed in the atmospheres and thus a vertically constant profile in the retrieval is physically realistic. N$_2$ is a filler gas in our atmosphere that adds to the other atmospheric constituents to ensure the atmospheric surface pressure totals to 2 bar. The retrieval is unable to directly constrain the abundance of N$_2$ in all cases due to the lack of spectral features in the IR, consistent with previous related works \citep[e.g.,][]{konrad2022LIFEIII,alei2022LIFEV}. 

Our CO$_2$ abundance is 10\% of the atmosphere at the surface, or 0.2 bar, in all models. The abundance of CO$_2$ can be retrieved from the primary IR CO$_2$ feature at 15 $\mu$m and the additional CO$_2$ features from 9-11 $\mu$m that appear in high CO$_2$ atmospheres. The retrieval is within $\pm$ 1 dex of the true abundance for all runs except the 0.5 bar H$_2$O vertically constant profile case. We note that the retrieval overestimates the CO$_2$ abundance consistently, ranging from 0.4 to 1 bar, in all models except the high-water altitude invariant cases. Starting at 10$^{-2}$ bar and higher water concentrations in the vertically constant profile models, the retrieved CO$_2$ concentration is notably lower, ranging from 0.08 bar down to 2$\times 10^{-5}$ bar due to increasing water absorption obscuring the CO$_2$ features. 

The retrieval assumes a vertically constant profile for all species, and so the retrieved posterior is always the same value for the surface and higher in the atmosphere. However, the water mid-IR spectral features are produced from a lower pressure than the surface, typically between the surface and 10$^{-1}$ bar. Thus, for the variable water profiles we would expect the retrieved value to be in between the simulation truth at the surface pressure of 2 bar and 10$^{-1}$ bar in the atmosphere. In Table \ref{table_LIFESIM} we list the simulation truth and retrieved posterior value for both the surface water concentration and the water concentration at 10$^{-1}$ bar to more accurately assess the retrieval results for the varying water profiles with altitude.

\begin{figure*}[ht!] 
\hspace{3cm} $\mathcal{M}_{\text{iso}}$ \hspace{4.7cm} $\mathcal{M}_{\oplus}$  \hspace{4.9cm}$\mathcal{M}_{\text{DPh}}$

\centering
\vspace{-0mm}
  \includegraphics[width=0.95\textwidth]{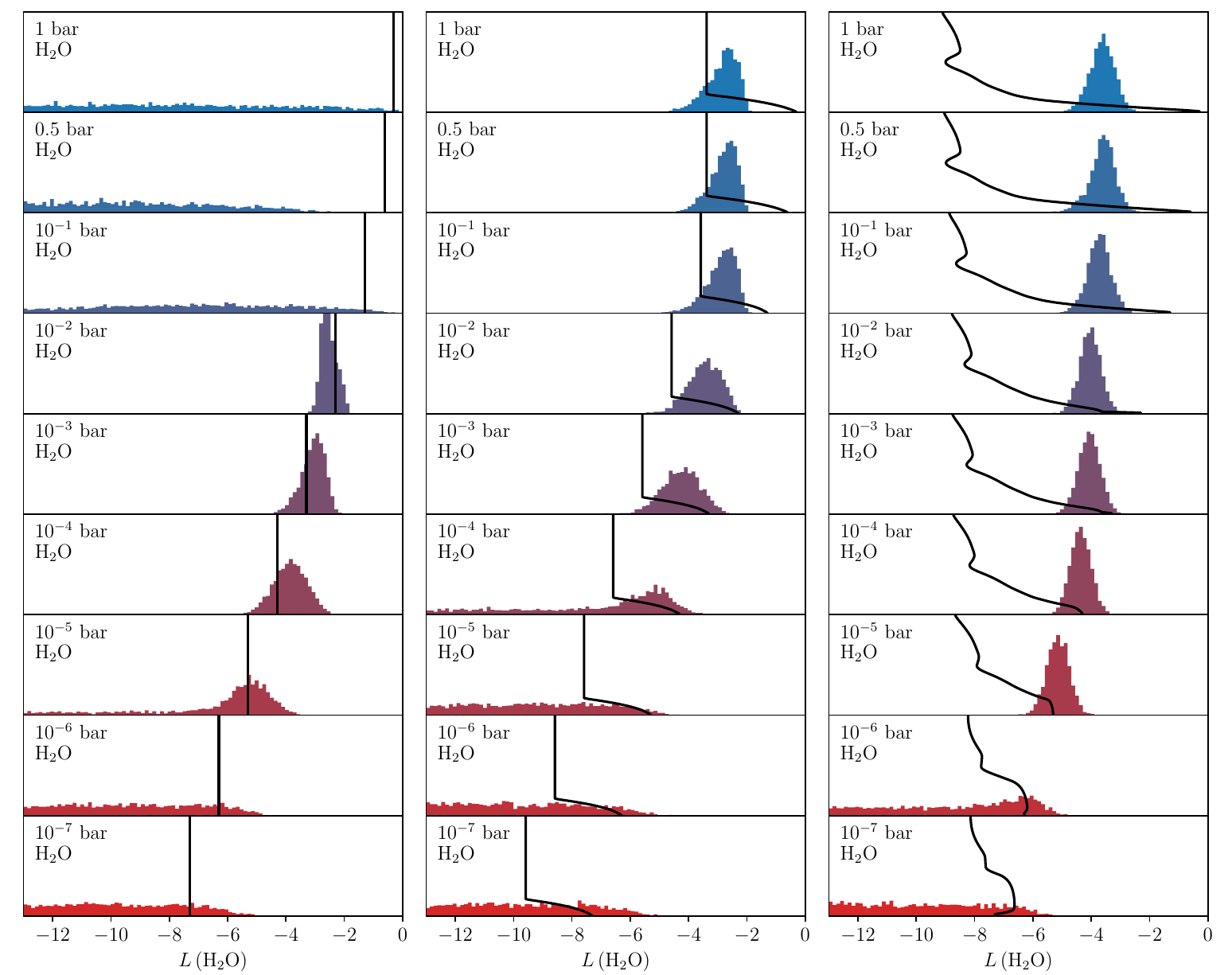}\caption{Water posteriors for simulations for a range of surface H$_2$O abundances from 1 bar to 10$^{-7}$ bar at a resolution and signal to noise ratio of R = 100 and S/N = 10, respectively. The black lines are the H$_2$O profiles in volume mixing ratios in log$_{10}$($L$) overlaying the H$_2$O posteriors. The left panel assumes an altitude invariant profile for water, commonly used in retrievals. The center panel fixes the surface abundance and then assumes a scaled Earth-like profile. The right panel fixes the surface H$_2$O abundance and has diffusion and photochemical production at all other altitudes. \label{posteriors}}
\end{figure*}

In Fig. \ref{posteriors} we show the water posteriors. The black lines are the water altitude profiles overlaying the water posteriors from the retrieval. The left panel assumes a vertically constant profile. The center panel fixes the surface abundance and then assumes a scaled Earth-like profile. The right panel fixes the surface water abundance and has diffusion and photochemical production at all other altitudes.

For the vertically constant profile altitude models at an R = 100 we obtain upper limits for the lowest water cases 10$^{-7}$ and 10$^{-6}$ bar; a sensitivity limit for 10$^{-5}$ bar with the retrieved value within 0.3 dex of the true value; a well-constrained model within 0.3-0.5 dex for the medium water cases 10$^{-4}$, 10$^{-3}$, and 10$^{-2}$ bar; and unconstrained for the high water cases of 10$^{-1}$, 0.5, and 1 bar. For R = 50 the main difference is that the 10$^{-5}$ bar run has a less strong sensitivity limit and the retrieved value is worse at 1.2 dex from the true value compared to the 0.3 dex at R = 100.

For the Earth-like water profiles at an R = 100 we obtain upper limits for the three lowest water cases 10$^{-7}$,  10$^{-6}$, and 10$^{-5}$ bar. At 10$^{-4}$ bar we have a sensitivity limit with the retrieved value falling between the surface and 0.1 bar true concentration. At 10$^{-3}$ bar and higher the priors are constrained and the retrieved values are between the surface and 0.1 bar H$_2$O concentration. For R = 50 the main difference is that the 10$^{-4}$ bar run has a less strong sensitivity limit and the retrieved value is 0.8 dex from the true value at 0.1 bar compared with 0.2 dex at R = 100.

For the diffusion and photochemistry profiles at an R=100 we obtain an upper limit for the two lowest water cases of 10$^{-7}$ and 10$^{-6}$ bar. For 10$^{-5}$ bar and higher the posteriors are constrained and the retrieved values are between the surface and 0.1 bar H$_2$O concentration. For R = 50 the main difference is that the 10$^{-6}$ bar run is unconstrained.

\onecolumngrid
\begin{longtable}{c l c r r r c c c}
\caption{Posterior retrieval results for H$_2$O models.}\label{table_LIFESIM}\\
     \toprule
    Surface H$_2$O & \colhead{$\theta_{\mathcal{M}}$} & Simulation & \multicolumn{3}{c}{Retrieved Posteriors} & \multicolumn{3}{c}{Posterior Classification}\\ 
    \cmidrule(lr){4-6}  \cmidrule(lr){7-9}
    [bar] &  & Truths & \multicolumn{1}{c}{$\mathcal{M}_{\text{iso}}$} & \multicolumn{1}{c}{$\mathcal{M}_{\oplus}$} & \multicolumn{1}{c}{$\mathcal{M}_{\text{DPh}}$} &\multicolumn{1}{c}{$\mathcal{M}_{\text{iso}}$} & \multicolumn{1}{c}{$\mathcal{M}_{\oplus}$} & \multicolumn{1}{c}{$\mathcal{M}_{\text{DPh}}$} \\
    \midrule
    \endfirsthead

    \multicolumn{9}{c}{\textit{Table continued from previous page}} \\
    \toprule
    Surface H$_2$O & \colhead{$\theta_{\mathcal{M}}$} & Simulation & \multicolumn{3}{c}{Retrieved Posteriors} & \multicolumn{3}{c}{Posterior Classification} \\
    \cmidrule(lr){4-6} \cmidrule(lr){7-9}
    (bar) &  & Truths & \multicolumn{1}{c}{$\mathcal{M}_{\text{iso}}$} & \multicolumn{1}{c}{$\mathcal{M}_{\oplus}$} & \multicolumn{1}{c}{$\mathcal{M}_{\text{DPh}}$} & \multicolumn{1}{c}{$\mathcal{M}_{\text{iso}}$} & \multicolumn{1}{c}{$\mathcal{M}_{\oplus}$} & \multicolumn{1}{c}{$\mathcal{M}_{\text{DPh}}$} \\
    \midrule
    \endhead

    \endfoot
    \bottomrule
    \endlastfoot

\multirow{6}{*}{10$^{-7}$} & $L(\text{P}_\text{surf})$ & 0.3 & $0.0^{+0.2}_{-0.2}$ \cmark & $0.0^{+0.2}_{-0.2}$ \cmark & $0.1^{+0.2}_{-0.2}$ \cmark & \statusbox{green}{C} &\statusbox{green}{C} &\statusbox{green}{C}\\
& $\text{T}_\text{surf}$ & 290 & $286^{+6}_{-6}$ \cmark & $287^{+6}_{-6}$ \cmark & $285^{+4}_{-4}$ \cmark & \statusbox{green}{C} & \statusbox{green}{C} & \statusbox{green}{C}\\
& $L(\text{H}_2\text{O})_\text{iso}$ & [-7.3, -7.3] & $-10.2^{+2.9}_{-2.8}$ \xmark & & & \statusbox{orange}{UL} &  & \\
& $L(\text{H}_2\text{O})_\oplus$ & [-7.3, -9.6] & & $-10.2^{+2.8}_{-2.8}$ \cmark & & \statusbox{white}{UL} & \statusbox{orange}{UL} & \statusbox{white}{UL}\\
& $L(\text{H}_2\text{O})_\text{ DPh}$ & [-7.3, -6.7] & & & $-10.7^{+3.0}_{-2.8}$ \xmark &  \statusbox{white}{UL} & \statusbox{white}{UL} & \statusbox{orange}{UL}\\
\midrule

 \multirow{6}{*}{10$^{-6}$} & $L(\text{P}_\text{surf})$ & 0.3 & $0.1^{+0.2}_{-0.2}$ \cmark & $0.0^{+0.2}_{-0.2}$ \cmark & $0.1^{+0.2}_{-0.2}$ \cmark &  \statusbox{green}{C} &\statusbox{green}{C} &\statusbox{green}{C}\\
& $\text{T}_\text{surf}$ & 290 & $284^{+6}_{-6}$ \cmark & $287^{+6}_{-7}$ \cmark & $284^{+5}_{-4}$ \cmark & \statusbox{green}{C} & \statusbox{green}{C} &\statusbox{green}{C}\\
& $L(\text{H}_2\text{O})\text{iso}$ & [-6.3, -6.3] & $-9.9^{+3.1}_{-3.2}$ \xmark & & & \statusbox{orange}{UL} & \statusbox{white}{UL} & \statusbox{white}{SL}\\
& $L(\text{H}_2\text{O})_\oplus$ & [-6.3, -8.6] & & $-10.4^{+3.0}_{-2.8}$ \xmark & & \statusbox{white}{UL} & \statusbox{orange}{UL} & \statusbox{white}{SL}\\
& $L(\text{H}_2\text{O})\text{DPh}$ & [-6.3, -6.3] & & & $-9.2^{+2.7}_{-3.7}$ \xmark & \statusbox{white}{UL} & \statusbox{white}{UL} & \statusbox{orange}{UL}\\
\midrule

\multirow{6}{*}{10$^{-5}$} & $L(\text{P}_\text{surf})$ & 0.3 & $0.1^{+0.2}_{-0.2}$ \cmark & $0.0^{+0.2}_{-0.2}$ \cmark & $0.1^{+0.2}_{-0.2}$ \cmark & \statusbox{green}{C} & \statusbox{green}{C} & \statusbox{green}{C}\\
& $\text{T}_\text{surf}$ & 290 & $284^{+7}_{-7}$ \cmark & $285^{+6}_{-6}$ \cmark & $285^{+5}_{-5}$ \cmark & \statusbox{green}{C} & \statusbox{green}{C} & \statusbox{green}{C}\\
& $L(\text{H}_2\text{O})\text{iso}$ & [-5.3, -5.3] & $-5.6^{+0.9}_{-4.0}$ \cmark & & & \statusbox{lime}{SL} & \statusbox{white}{UL} & \statusbox{white}{C}\\
& $L(\text{H}_2\text{O})_\oplus$ & [-5.3, -7.6] & & $-9.8^{+2.9}_{-3.2}$ \xmark & & \statusbox{white}{SL} & \statusbox{orange}{UL} & \statusbox{white}{C}\\
& $L(\text{H}_2\text{O})\text{DPh}$ & [-5.3, -6.1] & & & $-5.4^{+0.4}_{-0.4}$ \cmark & \statusbox{white}{SL} & \statusbox{white}{UL} & \statusbox{green}{C}\\
\midrule

\multirow{6}{*}{10$^{-4}$} & $L(\text{P}_\text{surf})$ & 0.3 & $0.1^{+0.2}_{-0.2}$ \cmark & $0.0^{+0.2}_{-0.2}$ \cmark & $0.1^{+0.2}_{-0.2}$ \cmark & \statusbox{green}{C} & \statusbox{green}{C} & \statusbox{green}{C}\\
& $\text{T}_\text{surf}$ & 290 & $286^{+7}_{-7}$ \cmark & $283^{+6}_{-6}$ \cmark & $285^{+5}_{-5}$ \cmark & \statusbox{green}{C} & \statusbox{green}{C} & \statusbox{green}{C}\\
& $L(\text{H}_2\text{O})\text{iso}$ & [-4.3, -4.3] & $-3.8^{+0.6}_{-0.6}$ \cmark & & & \statusbox{green}{C} & \statusbox{white}{SL} & \statusbox{white}{C}\\
& $L(\text{H}_2\text{O})_\oplus$ & [-4.3, -6.6] & & $-6.4^{+1.5}_{-5.3}$ \cmark & & \statusbox{white}{C} & \statusbox{lime}{SL} & \statusbox{white}{C}\\
& $L(\text{H}_2\text{O})\text{DPh}$ & [-4.3, -6.4] & & & $-4.6^{+0.3}_{-0.3}$ \cmark & \statusbox{white}{C} & \statusbox{white}{SL} & \statusbox{green}{C} \\
\midrule

\multirow{6}{*}{10$^{-3}$} & $L(\text{P}_\text{surf})$ & 0.3 & $0.0^{+0.2}_{-0.2}$ \cmark & $0.0^{+0.2}_{-0.2}$ \cmark & $0.2^{+0.2}_{-0.2}$ \cmark & \statusbox{green}{C} & \statusbox{green}{C} & \statusbox{green}{C}\\
& $\text{T}_\text{surf}$ & 290 & $286^{+6}_{-6}$ \cmark & $285^{+7}_{-7}$ \cmark & $285^{+5}_{-5}$ \cmark & \statusbox{green}{C} & \statusbox{green}{C} & \statusbox{green}{C}\\

& $L(\text{H}_2\text{O})_\text{iso}$ & [-3.3, -3.3] & $-3.0^{+0.4}_{-0.5}$ \cmark & & & \statusbox{green}{C} & \statusbox{white}{C} & \statusbox{white}{C}\\
& $L(\text{H}_2\text{O})_\oplus$ & [-3.3, -5.6] & & $-4.3^{+0.6}_{-0.7}$ \cmark & & \statusbox{white}{C} & \statusbox{green}{C} & \statusbox{white}{C}\\
& $L(\text{H}_2\text{O})_\text{DPh}$ & [-3.3, -6.4] & & & $-4.3^{+0.4}_{-0.4}$ \cmark & \statusbox{white}{C} & \statusbox{white}{C} & \statusbox{green}{C}\\
\midrule

\multirow{6}{*}{10$^{-2}$} & $L(\text{P}_\text{surf})$ & 0.3 & $0.6^{+0.3}_{-0.4}$ \cmark& $0.0^{+0.2}_{-0.2}$ \cmark& $0.1^{+0.2}_{-0.2}$ \cmark & \statusbox{green}{C} & \statusbox{green}{C} & \statusbox{green}{C}\\
& $\text{T}_\text{surf}$ & 290 & $302^{+14}_{-13}$ \cmark& $283^{+7}_{-7}$ \cmark & $283^{+5}_{-5}$ \cmark & \statusbox{green}{C} & \statusbox{green}{C} & \statusbox{green}{C}\\

& $L(\text{H}_2\text{O})\text{iso}$ & [-2.3, -2.3] & $-2.6^{+0.3}_{-0.2}$ \cmark & & & \statusbox{green}{C} & \statusbox{white}{C} & \statusbox{white}{C} \\
& $L(\text{H}_2\text{O})_\oplus$ & [-2.3, -4.6] & & $-3.4^{+0.6}_{-0.6}$ \cmark & & \statusbox{white}{C} & \statusbox{green}{C} & \statusbox{white}{C}\\
& $L(\text{H}_2\text{O})\text{DPh}$ & [-2.3, -6.5] & & & $-4.3^{+0.3}_{-0.3}$ \cmark & \statusbox{white}{C} & \statusbox{white}{C} & \statusbox{green}{C}\\
\midrule

\multirow{6}{*}{10$^{-1}$} & $L(\text{P}_\text{surf})$ & 0.3 & $-2.5^{+0.3}_{-0.3}$ \xmark & $-0.1^{+0.3}_{-0.2}$ \cmark & $0.0^{+0.2}_{-0.2}$ \cmark & \statusbox{yellow1}{C} & \statusbox{green}{C} & \statusbox{green}{C}\\
& $\text{T}_\text{surf}$ & 290 & $183^{+1}_{-1}$ \xmark & $270^{+6}_{-6}$ \cmark & $275^{+5}_{-5}$ \cmark & \statusbox{yellow1}{C} & \statusbox{green}{C} & \statusbox{green}{C}\\

& $L(\text{H}_2\text{O})\text{iso}$ & [-1.3, -1.3] & $-7.2^{+3.7}_{-4.1}$ \xmark & & & \statusbox{red}{UC} & \statusbox{white}{C} & \statusbox{white}{C} \\
& $L(\text{H}_2\text{O})_\oplus$ & [-1.3, -3.6] & & $-2.8^{+0.4}_{-0.6}$ \cmark & & \statusbox{white}{UC} & \statusbox{green}{C} & \statusbox{white}{C}  \\
& $L(\text{H}_2\text{O})\text{DPh}$ & [-1.3, -6.7] & & & $-4.0^{+0.4}_{-0.4}$ \cmark & \statusbox{white}{UC} & \statusbox{white}{C} & \statusbox{green}{C} \\
\midrule

\multirow{6}{*}{0.5} & $L(\text{P}_\text{surf})$ & 0.3 & $-0.9^{+0.4}_{-0.3}$ \xmark & $-0.2^{+0.3}_{-0.2}$ \cmark & $-0.1^{+0.2}_{-0.2}$ \cmark & \statusbox{yellow1}{C} & \statusbox{green}{C} & \statusbox{green}{C}\\
& $\text{T}_\text{surf}$ & 290 & $182^{+1}_{-1}$ \xmark & $260^{+6}_{-5}$ \xmark & $268^{+5}_{-4}$ \xmark & \statusbox{yellow1}{C} & \statusbox{yellow1}{C} & \statusbox{yellow1}{C} \\

& $L(\text{H}_2\text{O})\text{iso}$ & [-0.6, -0.6] & $-9.6^{+3.6}_{-3.3}$ \xmark & & & \statusbox{red}{UC} & \statusbox{white}{C} & \statusbox{white}{C} \\
& $L(\text{H}_2\text{O})_\oplus$ & [-0.6, -3.4] & & $-2.7^{+0.4}_{-0.5}$ \cmark & & \statusbox{white}{UC} & \statusbox{green}{C} & \statusbox{white}{C} \\
& $L(\text{H}_2\text{O})\text{DPh}$ & [-0.6, -7.1] & & & $-3.9^{+0.4}_{-0.4}$ \cmark & \statusbox{white}{UC} & \statusbox{white}{C} & \statusbox{green}{C} \\
\midrule
   
\multirow{6}{*}{1.0} & $L(\text{P}_\text{surf})$ & 0.3 & $-2.6^{+0.3}_{-0.3}$ \xmark & $-0.2^{+0.3}_{-0.2}$ \cmark & $-0.1^{+0.2}_{-0.2}$ \cmark & \statusbox{yellow1}{C} & \statusbox{lime}{SL} & \statusbox{green}{C}\\
& $\text{T}_\text{surf}$ & 290 & $183^{+1}_{-1}$ \xmark & $255^{+6}_{-5}$ \xmark & $266^{+5}_{-5}$ \xmark & \statusbox{yellow1}{C} & \statusbox{yellow1}{C} & \statusbox{yellow1}{C}\\
& $L(\text{H}_2\text{O})\text{iso}$ & [-0.3, -0.3] & $-8.1^{+4.7}_{-4.1}$ \xmark & & & \statusbox{red}{UC} & \statusbox{white}{C} & \statusbox{white}{C} \\
& $L(\text{H}_2\text{O})_\oplus$ & [-0.3, -3.4] & & $-2.8^{+0.4}_{-0.7}$ \cmark & & \statusbox{white}{UC} & \statusbox{green}{C} & \statusbox{white}{C} \\
& $L(\text{H}_2\text{O})\text{DPh}$ & [-0.3, -7.1] & & & $-3.9^{+0.4}_{-0.4}$ \cmark & \statusbox{white}{UC} & \statusbox{white}{C} & \statusbox{green}{C} \\

\end{longtable}
\vspace{-0.5\baselineskip} 
\hspace*{0.7cm}
\begin{minipage}{\dimexpr\linewidth-2cm} 
\footnotesize
\textbf{Notes.} Summary of the retrieval results for a LIFE mission architecture for temperature (K), pressure (bar), and H$_2$O abundance of a 2 bar prebiotic terrestrial atmosphere with surface concentrations of H$_2$O ranging from 10$^{-7}$ to 1 bar. $L$ denotes log$_{10}$. We provide the simulation truth at the surface and at 10$^{-1}$ bar in brackets ([$\nu_\text{{2 bar}}$, $\nu_{10^{-1} \text{ bar}}$]). We model three water profiles based on a constant H$_2$O concentration with height ($\mathcal{M}_{\text{iso}}$), an Earth-like profile ($\mathcal{M}_{\oplus}$), and a profile considering only diffusion and photochemical production of H$_2$O ($\mathcal{M}_{\text{DPh}}$). All abundances and pressures are given in volume mixing ratios in log$_{10}$. We provide 1$\sigma$ confidence intervals (the 16th and 84th percentile of the logistic function) as denoted by the $\pm$ in the retrieved posteriors. We consider a retrieved posterior to be accurate if it is within $\pm$0.5 dex of the true pressure, $\pm$20 K of the true surface temperature, and $\pm$1 dex of the water abundance following \citet{konrad2022LIFEIII}.  We show the posterior classification for each simulation as denoted by: \statusbox{green}{C} = constrained; \statusbox{yellow1}{C} = constrained but inaccurate; \statusbox{lime}{SL} = sensitivity limit; \statusbox{orange}{UL} = upper limit, and \statusbox{red}{UC} = unconstrained.
\end{minipage}

\par\medskip 
\twocolumngrid

\section{Discussion} \label{sec:discussion}

Although retrieval models typically assume vertically constant profiles for atmospheric gases, physics and chemistry based forward models indicate that for most gases the vertical abundance will vary due to photochemistry, photolysis, diffusion, and atmospheric escape. Progress is being made to improve on this issue with retrievals \citep[see][]{bourrier2020,rowland2023,konrad2024,nasedkin2025}), and this paper adds to the literature to indicate the importance of the vertical profile assumptions on retrieving water in terrestrial planets. On Earth, water is confined primarily to the troposphere. As air rises and cools the water condenses and precipitates out, leaving trace amounts in the stratosphere on temperate Earth-like planets. If the cold trap is strong it will inhibit upward mixing. As a planet warms and approaches a moist or runaway greenhouse effect, evaporation will increase and the water approaches a vertically constant altitude profile (see e.g. Fig. 1 in \citet{kasting2015} and Fig. 2 in \citet{kastingHZ1993}). Once there is no cold trap to keep water in the tropopause and water is lofted into the stratosphere, it is then subject to photolysis followed by hydrogen escape, and the planet undergoes permanent water loss. An abiotic oxygen rich atmosphere may follow \citep{ingersoll1969, kasting1988, wordsworth2014, luger2015, harman2015} depending on the redox state of the planet surface \citep{schaefer2015} and whether or not the atmosphere is subject to combustion \citep{grenfell2018}. 

The precise detectability of water is sensitive to the profile assumed. For vertically constant water profiles, there is a Goldilocks region of water detectability. If there is too much, water swamps out its own signal. If there is too little, there is not enough water to detect beyond an upper limit. This is because at high water abundances the strong H$_2$O line opacity and continuum absorption render the atmosphere optically thick over much of the spectrum, so the emergent radiation originates from higher, cooler layers and becomes insensitive to the deeper H$_2$O abundance. Values an order of magnitude above or below modern Earth's are detectable. However, vertically constant profiles are not the likely water profile for temperate Earth-like planets. For models with physically motivated water profiles based on modern Earth with a large surface reservoir or profiles based on evaporation, diffusion, and photochemistry from a fixed surface concentration of water we find that water abundance is detectable for a wider range of values. For Earth-like water profiles we can detect water two orders of magnitude above and below Earth's. For a profile based on diffusion and photochemistry, we can detect water two orders of magnitude above and four orders of magnitude below modern Earth.  For all water profiles we can detect around Earth surface water concentrations and for no model can we detect Mars' concentrations. For reference, Earth has a volume mixing ratio surface water abundance of $\sim 10^{-2}$ and Mars has one of $\sim 10^{-6}$.

\citet{konrad2024} uses a complementary approach to test for a modern Earth atmosphere water profiles retrieved with the LIFE mission by including a condensation model in the troposphere and varying the stratospheric water abundance by a drying parameter. They find that the simplifying assumptions of constant water profiles and cloud free atmospheres biases the P-T structure and atmospheric composition, though the Bayesian retrieval does not show a preference for more complex models. Yet, because forward modeling allows us to impose physically-informed constraints, it will yield more reliable retrieved atmospheric values. We find additionally by varying the water abundance and adopting two different vertically-varying profiles that the posteriors are much more constrained and the retrieved values are more accurate once the water abundance is different from modern Earth. 

Furthermore, in the case of the high-water vertically constant profiles the retrieval is confidently wrong, retrieving small error bars and largely classifying the posterior as constrained. These high-water cases hide themselves, mimicking a planet that is much colder and with a lower pressure atmosphere. This is due to strong water opacity swamping out other spectral features so completely that a low pressure atmosphere with minimal CO$_2$ is preferred instead of a 2 bar water-rich world. This is a bias where the data are degenerate with two models and it would be difficult to discriminate between the two scenarios with these future missions as planned.

Detectable atmospheric water vapor may itself be indicative of a surface ocean indirectly since extremely limited water vapor would not be stable due to its reactivity with surface minerals; it would be incorporated into the rock and precipitate out of the atmosphere (Kasting, personal communication). For extremely high amounts of water vapor, such as in a moist or run-away greenhouse, the atmosphere would be opaque and water would also not be constrained, and the planet also wouldn't be habitable \citep{Boer2025ApJ}. Thus the Goldilocks region of not too much and not too little water is the region where water is detectable and indicates of habitable conditions. 

There may be additional ways to determine the lack of water. An ocean's thermal inertia would reduce the thermal gradient between the day and night-side, and thus the absence of a pronounced horizontal temperature gradient might be observable through thermal-emission direct imaging and indicate the lack of a surface ocean \citep{fujii2025}. For M dwarf planets, the detection of sulfur dioxide may also indicate a lack of an ocean. SO$_2$ would otherwise rain out in a planet with a hydrological cycle \citep{loftus2019}. Around Sun-like stars, SO$_2$ is efficiently destroyed by photochemistry, however, in the low near-UV environment of an M dwarf host star, SO$_2$ can reach detectable levels, particularly for the feature at 7-8 $\mu$m, enabling the inference of a lack of water oceans \citep{jordan2025}.

The stellar spectral type will change the water profile due to the UV environment. Stratospheric water from methane oxidation increases when there is a higher UV environment because OH, which drives the reaction, is produced by UV \citep{grenfell2007}. However, there is a balance as H$_2$O is also destroyed by UVC photons in the far-UV region and, in particular, at the Lyman-$\alpha$ wavelength which is highest for M dwarf stars \citep{miguel2015,gebauer2018, rugheimer2015b}. When increasing the UV by moving the Earth-Sun system inward, the additional stratospheric H$_2$O produced by CH$_4$ oxidation has been shown to have a negligible impact on the H$_2$O feature across 4.0-8.0 $\mathrm{\mu}$m. However, the additional greenhouse heating does increase the thermal broadening of the CO$_2$ fundamental band between 12.0-16.0 $\mathrm{\mu}$m \citep{taysum2024}. We modeled only the Earth-Sun system here and future work will determine the applicability of these results to other star types. 

While not all planets transit, for those that do the HZ inner edge discontinuity may provide a statistical framework to better understand exoplanet populations at the HZ inner edge. \citet{schlecker2024} posits that there might be a demographic imprint of the transition of habitable to uninhabitable planets at the inner edge of the HZ if some fraction are going through a runaway greenhouse process. This potential radius-density discontinuity at the inner edge of the HZ would be detectable with PLATO if 10\% of planets have an inflated atmosphere and thus a higher radius in transit during a runaway greenhouse. Such planets cooling down from their primordial heat during an early runaway phase would be measurable with LIFE, potentially allowing constraints on the water abundance on young Earth-sized exoplanets \citep{Cesario2024AA}. These measurements would not constrain the water abundances of an individual, temperate exoplanet but provide contextual information on the overall planet population at the inner edge of the habitable zone and the initial water abundances terrestrial exoplanets are born with. The LIFE mission is not directly measuring the radius of the planet in transit, and is fitting the temperature and radius from the blackbody spectrum to the extracted flux. Our retrieval model retrieves a lower radius and lower temperature for those planets with extremely water rich atmospheres. If the planet transits and we are able to get a precise mass, radius and density, this added information may improve the retrieved posteriors with the LIFE mission. 

By not directly modeling clouds on the spectra, these results are likely optimistic. Clouds decrease the overall emitted flux of an Earth-like planet in the IR because they radiate at lower temperatures and therefore decrease the equivalent width of absorption features \citep[see e.g.][]{kaltenegger2007, kitzmann2011a}. They can in some cases increase the relative depth of a spectral feature due to lowering the continuum temperature of the planet \citep[see e.g.][]{rugheimer2013}. Clouds can block access to the surface, particularly in the visible \citep{kitzmann2011b,kawashima2019}, but also in the IR \citep{vasquez2013b}. There is a degeneracy between a cool, cloud-less planet and a habitable one with high-altitude clouds or even an exo-Venus \citep{selsis2000,barstow2016,lustig2019, Konrad2023}.

\section{Conclusions} \label{sec:conclusions}

In this study, we assess the ability of LIFE to retrieve the water abundance to confirm surface habitability in terrestrial HZ exoplanets. We modeled three vertical H$_2$O abundance profiles with 9 surface H$_2$O concentrations and tested how they impacted the retrieved posteriors with the LIFE mission at an R = 50, 100 and a reference S/N = 10 at 11.2 $\mu$m. Our water profiles were modeled by (1) a constant vertical mixing ratio profile, (2) a Manabe-Wetherald based Earth-like profile with a fixed surface mixing ratio, and (3) a profile where the surface concentration of water is fixed and water in the layers above the surface is determined by only diffusion and photochemical production. 

Retrievals often assume a vertically constant abundance profiles of atmospheric species to reduce computational complexity, yet increasingly there a growing research showing this may bias the retrieved posteriors \citep{rowland2023, konrad2024, nasedkin2025}, and this work demonstrates the variation in the detectability due to vertical water profile. 

We find the original minimum requirements of spectral resolution and S/N for LIFE (R = 50, S/N = 10)\footnote{See online repository for R = 50 and R=100 results: \url{https://github.com/LIFE-SpaceMission/Detecting_Oceans}} determined in \citet{konrad2022LIFEIII} remain sufficient for detecting Earth-like water concentrations in all models. However, an R = 100 would allow for a detection of H$_2$O at concentrations an order of magnitude lower than that of an R = 50, indicating that a higher resolution would allow for a tighter constraint on habitable conditions. Additionally, \citet{konrad2024} shows that an R = 100 is necessary to for a bias-free detection of CH$_4$ at Earth-like concentrations.

If a planet has constant water abundance with height as is expected in a runaway greenhouse, at  concentrations greater than 0.1 bar H$_2$O would not be detectable. Water swamps out its own signal and the spectrum is degenerate with a cooler planet radiating as a blackbody.  Water becomes increasingly closer to a constant mixing ratio with height as an Earth-like planet approaches a run-away greenhouse phase as can be seen in Fig 5 of \citet{kasting1988}, but is sensitively affected by the geochemistry of the underlying interior \citep{Boer2025ApJ}. Modeling a full self-consistent run-away greenhouse atmosphere is beyond the scope of this manuscript and will be tackled in future works. 

LIFE would be unable to directly detect surface water via a reflectance signal such as ocean glint, however its sensitivity to water vapor and other molecules which absorb in the mid-IR allow it to more completely characterize a planet by putting constraints on the oxidizing and reducing species in the atmosphere. 

Our models indicate that water vapor at Mars' concentrations are not detectable and we only retrieve an upper limit. We can accurately constrain water for abundances a few orders of magnitude greater and less than modern Earth, profile depending. Extremely high water atmospheres and water poor atmospheres are unconstrained or have only an upper limit. However, detecting water vapor may be all that is required to confirm a habitable surface \citep{desmarais2002}. This is due to the high reactivity of water with rock and surface minerals, meaning that in a water poor planet, water would quickly precipitate out of the atmosphere and be buried in the rocks (Kasting, personal communication). For the highest water atmospheres where H$_2$O is blocking its own features, those planets are also likely not habitable due to being in the midst of a runaway greenhouse process. Thus, detectable levels of water vapor in an atmosphere may also indicate the presence of surface liquid water and habitability.

\section*{Acknowledgments}
We thank Jim Kasting for his helpful comments on the the link between surface water and water vapour. We thank Felix Dannert for his helpful comments on instrument systematics in LIFE. SR acknowledges NSERC grant RGPIN-2022-04588. Part of this work has been carried out within the framework of the National Centre of Competence in Research PlanetS supported by the Swiss National Science Foundation under grants 51NF40\_182901 and 51NF40\_205606. EA and SPQ acknowledge the financial support of the SNSF. EA’s research was partly supported by an appointment to the NASA Postdoctoral Program at the NASA Goddard Space Research, administered by Oak Ridge Associated Universities under contract with NASA (ORAU-80HQTR21CA005). TL was supported by the Branco Weiss Foundation, the Netherlands eScience Center (PROTEUS project, NLESC.\-OEC.\-2023.\-017), the Alfred P. Sloan Foundation (AEThER project, G-2025-25284), NASA’s Nexus for Exoplanet System Science research coordination network (Alien Earths project, 80NSSC21K0593), and the Dutch Research Council NWA-ORC PRELIFE Consortium (NWA.1630.23.013). D.K. acknowledges the support from the Swiss National Science Foundation under the grant 200021-231596.



\section*{References}
\expandafter\ifx\csname natexlab\endcsname\relax\def\natexlab#1{#1}\fi
\providecommand{\url}[1]{\href{#1}{#1}}
\providecommand{\dodoi}[1]{doi:~\href{http://doi.org/#1}{\nolinkurl{#1}}}
\providecommand{\doeprint}[1]{\href{http://ascl.net/#1}{\nolinkurl{http://ascl.net/#1}}}
\providecommand{\doarXiv}[1]{\href{https://arxiv.org/abs/#1}{\nolinkurl{https://arxiv.org/abs/#1}}}

\clearpage

\appendix 
\section{Selected retrieval corner plots} \label{sec:appendix}

\begin{figure*}[h]
    \centering
    \includegraphics[width=0.94\textwidth]{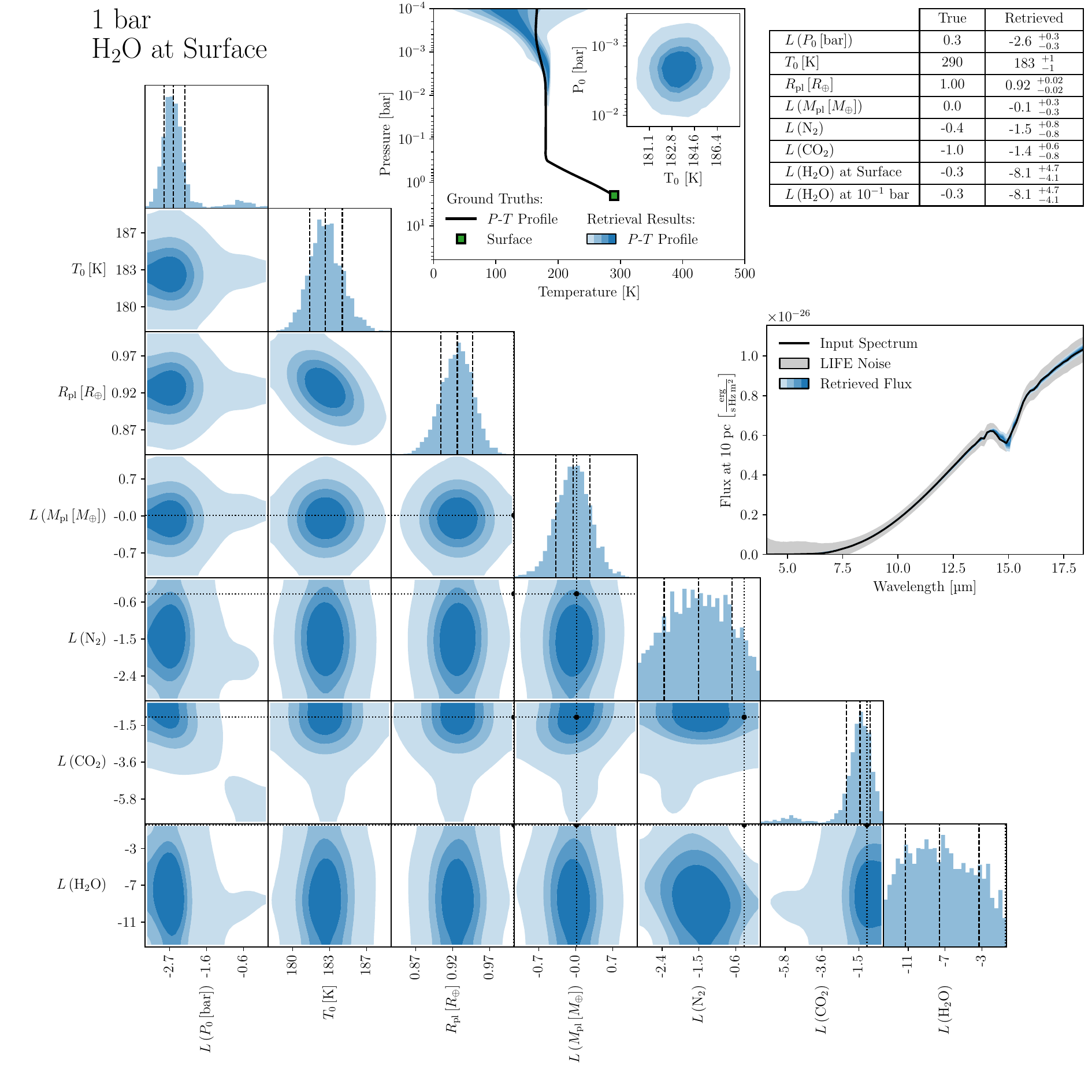}
    \caption{Corner plot for the posterior distributions from the retrievals of a 1 bar surface H$_2$O vertically constant profile model. Black dotted lines indicate the expected values, abundances reported in volume mixing ratios. Retrieved values are the dashed black lines (both plotted as the median and 1$\sigma$ uncertainties) and are in the table with the expected values. $L$ denotes log$_{10}$. Top-middle: The P–T profile with 2D histogram. Middle-right: input vs retrieved spectra with gray-shading showing \textsc{LIFEsim} uncertainty (S/N= 10). Shaded contours represent 1–4$\sigma$ confidence levels.\label{firstcornerplot}}
\end{figure*}

\begin{figure*}[h]
    \centering
    \includegraphics[width=0.94\textwidth]{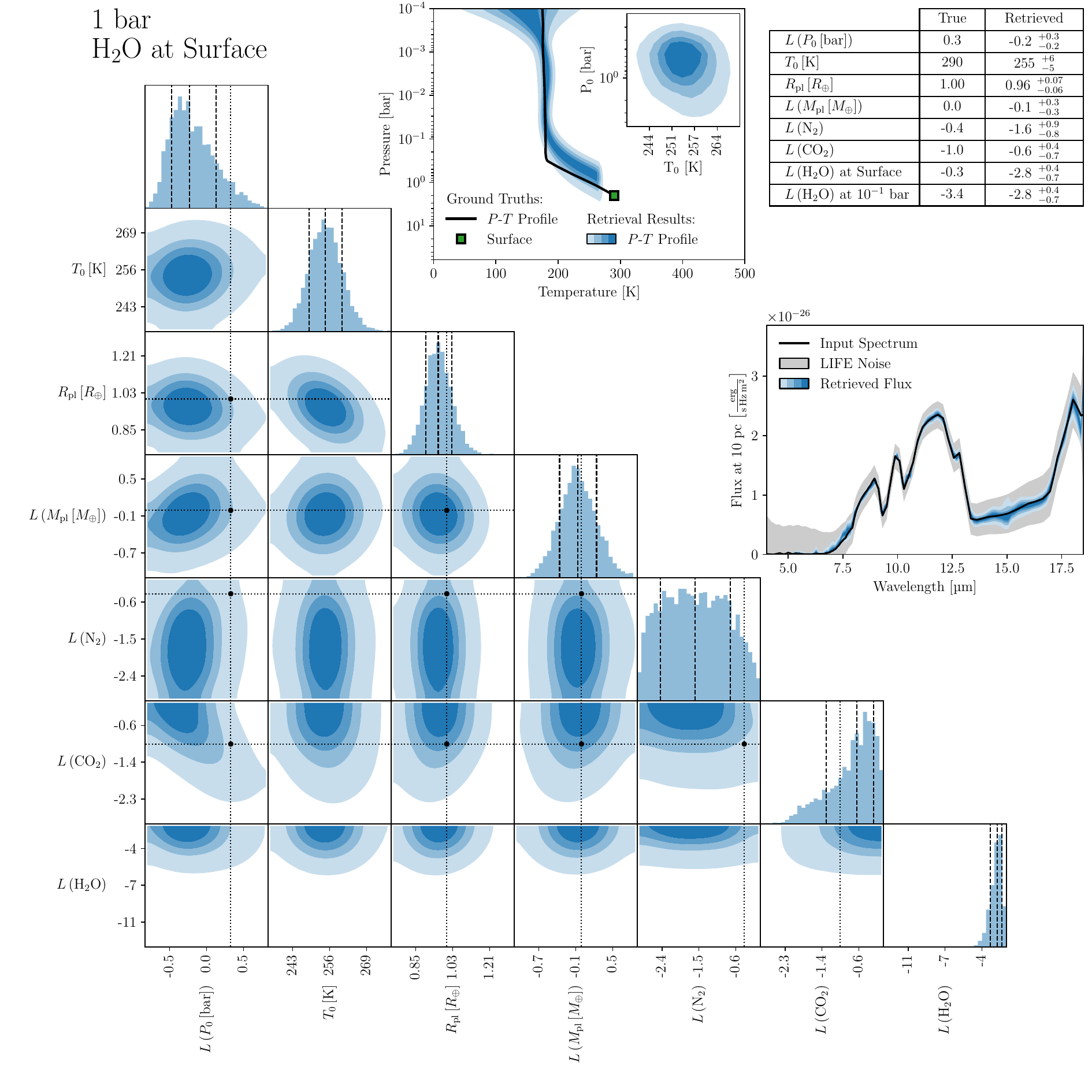}
    \caption{As for Figure \ref{firstcornerplot} except for the 1 bar surface H$_2$O Earth-like water profile model.}
\end{figure*}

\begin{figure*}[h]
    \centering
\includegraphics[width=0.94\textwidth]{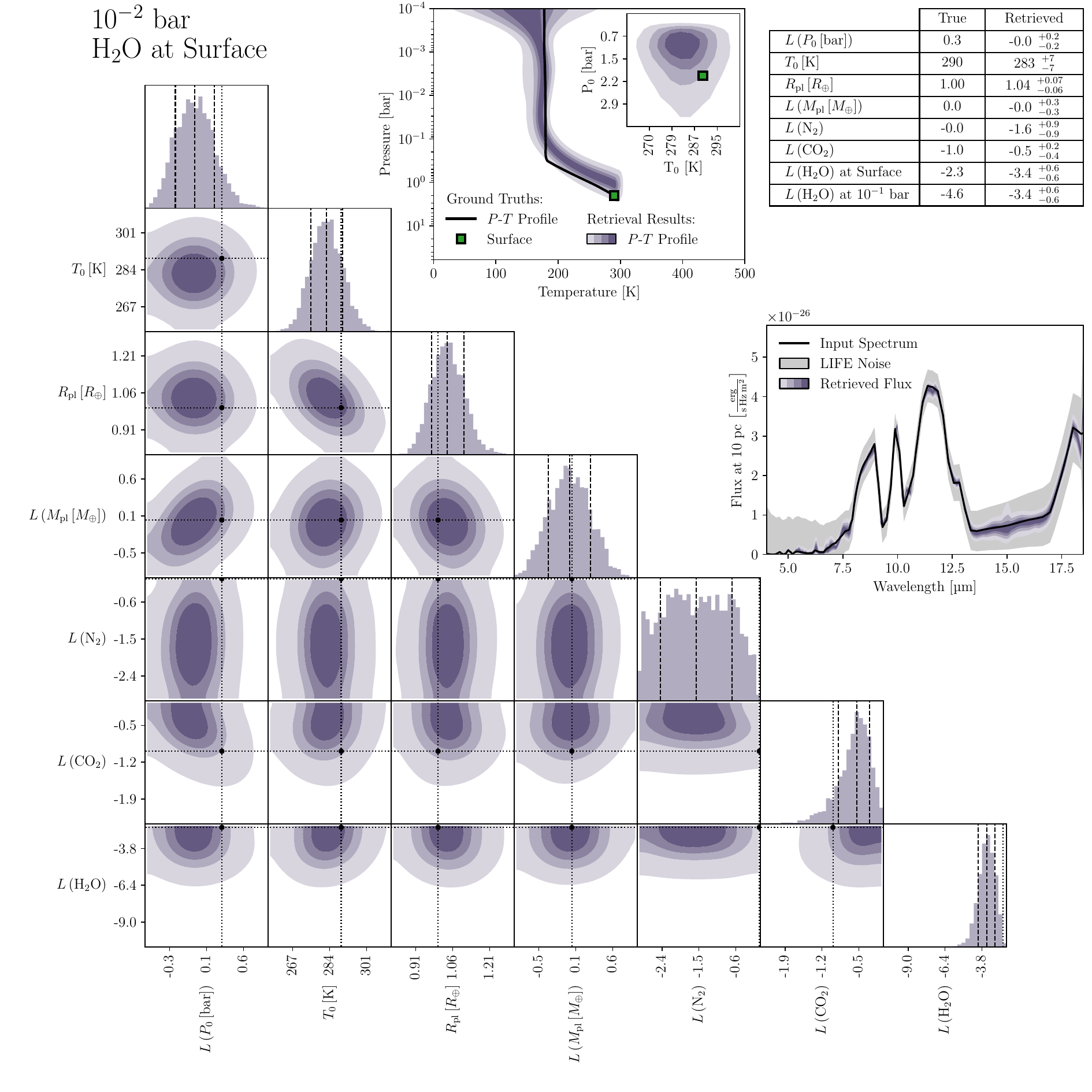}
    \caption{As for Figure \ref{firstcornerplot} except for the 10$^{-2}$ bar surface H$_2$O Earth-like water profile model. Note this also corresponds to a surface water concentration similar to modern Earth.}
\end{figure*}

\begin{figure*}[h]
    \centering
\includegraphics[width=0.94\textwidth]{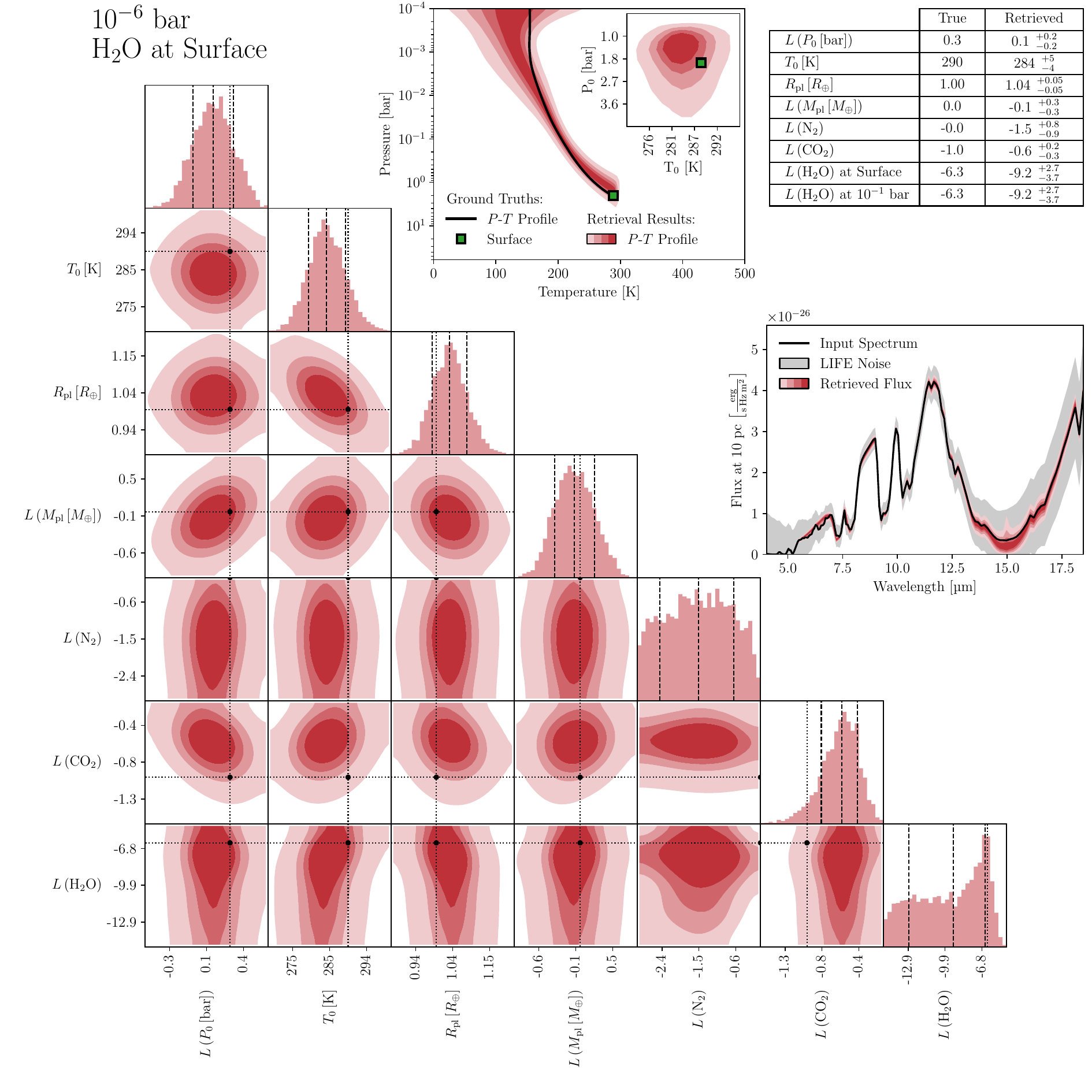}
    \caption{As for Figure \ref{firstcornerplot} except for the 10$^{-6}$ bar surface H$_2$O with diffusion and photochemistry only contributing to the H$_2$O abundance above the surface layer. At these low surface H$_2$O concentrations, stratospheric H$_2$O is generated primarily by methane oxidation photochemistry detailed in equation \ref{methaneoxidation}. The posterior classification provides only an upper limit for H$_2$O abundance at this low concentration. Note this also corresponds to a surface water concentration similar to modern Mars.}
\end{figure*}

\end{document}